\documentclass[aps,prb,twocolumn]{revtex4}

\usepackage{graphicx}
\usepackage{dcolumn}
\usepackage{amsmath}

\begin{document}

\title{Polar phonons and spin excitations coupling in multiferroic BiFeO$_3$ crystals}

\author{P. Rovillain}
\author{M. Cazayous}
\author{Y. Gallais}
\author{A. Sacuto}
\affiliation{Laboratoire Mat\'eriaux et Ph\'enom\`enes Quantiques (UMR 7162 CNRS), 
Universit\'e Paris Diderot-Paris 7, 75205 Paris cedex 13, France}

\author{R.P.S.M. Lobo}
\address{Laboratoire Photons et Mati\'ere (CNRS UPR 5), ESPCI, Universit\'e Pierre et Marie Curie, 10 rue Vauquelin, 75231 Paris Cedex 5, France}

\author{D. Lebeugle}
\author{D. Colson}
\affiliation{Service de Physique de l'Etat Condens\'e, DSM/DRECAM/SPEC, CEA Saclay, 91191 Gif-sur-Yvette, France}

\date{\today}
     
\begin{abstract}

Raman scattering measurements on BiFeO$_3$ single crystals show an important coupling between the magnetic order and lattice vibrations. The temperature evolution of phonons shows that the  lowest energy $E$ and $A_1$ phonon modes are coupled to the spin order up to the N\'eel temperature. Furthermore, low temperature anomalies associated with the spin re-orientation are observed simultaneously in both the $E$ phonon and the magnon. These results suggest that magnetostriction plays an important role in BiFeO$_3$.
\end{abstract}

\pacs{77.80.Bh, 75.50.Ee, 75.25.+z, 78.30.Hv}

\maketitle
The recent discovery of a new class of oxides exhibiting a strong magnetic and ferroelectric coupling with potential applications \cite{Scott} in spintronic devices \cite{Zhao, Cheong} greatly renewed the interest in the understanding of the fundamental excitations in multiferroic materials. \cite{Katsura, Sergienko, Mostovoy, Betouras}
Bismuth ferrite (BiFeO$_3$) belongs to this new family and is the only compound showing magnetoelectric coupling at room temperature. In magneto-electric multiferroics, spin waves (magnons) and electrical polarization waves induced by lattice displacements (phonon modes) are expected to be intimately connected to each other.
Such coupling gives rise to the so-called electromagnons --- spin waves that carry a electric dipole moment thus generating low frequency magneto-optical resonances in the dielectric susceptibility.

Electromagnons have been predicted by several models \cite{Maugin, Sousa, Cano} and have already been detected in the improper ferroelectric materials YMn$_2$O$_5$ and TbMn$_2$O$_5$.\cite{Pimenov, Sushkov} In these compounds, ferroelectricity is a consequence of the antiferromagnetic order and plausible microscopic magneto-electric coupling mechanisms include the inverse Dzyaloshinskii-Moriya (DM) interaction.\cite{Katsura, Sergienko}

The magneto-electric coupling in BiFeO$_3$ (BFO) is under intensive debate. This material is a proper ferroelectric exhibiting weak antiferromagnetism therefore making the inverse DM picture unlikely.\cite{Ederer} Recently, Sushkov {\it et al.} \cite{Sushkov2} proposed another coupling mechanism based on the isotropic Heisenberg exchange and magnetostrictive coupling of spins to a polar lattice mode. Several experiments have attempted to track the coupling mechanism between the dielectric and magnetic orders. Among them second-harmonic generation as well as dielectric and magnetic measurements are the most important.\cite{Fiebig,Kimura} However a simultaneous analysis of both magnetic and ferroelectric orders remains a challenge. Optical spectroscopies seem to be the ideal probes for carrying out such a study.  

The vibrational spectra of BiFeO$_3$ (BFO) is well characterized and only minor questions remain open. 
Raman \cite{Cazayous}, infrared \cite{Kamba, Lobo}, and lattice dynamical calculations \cite{Hermet, Tutuncu} are converging to a common phonon picture.
Infrared measurements reveal that the lowest frequency phonon mode at 75~cm$^{-1}$ has $E$ symmetry. 
This mode is responsible for the temperature dependence of the dielectric constant and is a potential candidate to represent the soft mode driving the ferroelectric transition.\cite{Kamba, Lobo} However, further phonon measurements, specially in the paraelectric phase are necessary to establish this view. 
Recent Raman experiments have detected low frequency magnons and suggest that their electrical dipole character advocates in favor of electromagnons.\cite{Cazayous1, Singh}
In this perspective, one of the key points in understanding the magnetoelectric coupling in BFO is to discover the phonon modes responsible for the magnons electrical activity.

Here, we present Raman measurements on BFO single crystals from 7~K to 1000~K. We show that the lowest $E$ phonon mode is related to the spin order in BFO. 
We detect two anomalies in both the intensities and frequencies of the magnons and this $E$ phonon near 130~K and 210~K. 
Theses anomalies come from a spin reorientation of the Fe$^{3+}$ magnetic moments out of the magnetic cycloidal plane of BFO.
At the N\'eel temperature, both the $E$ phonon and the lowest $A_1$ vibrational mode exhibit a clear change in their temperature dependence indicating their coupling to the spins.
Besides giving further evidence for an electromagnon picture in BFO, these simultaneous observations reveal that the ferroelastic deformation is an important feature in the spin-phonon coupling. This result strongly suggets that the magneto-electric coupling in BFO is mediated by magneto-striction and piezoelectric effects.

Millimeter-size single crystals of BFO shaped as dark red platelets were grown in air using a Bi$_2$O$_3$-Fe$_2$O$_3$ flux in a alumina crucible.\cite{Lebeugle}
At room temperature, BFO is a rhombohedrally distorted perovskite with space group $R3c$.
Below its Curie temperature, $T_C \sim 1100$~K, BFO has a large spontaneous electrical polarization ($\sim 60 \mu$C/cm$^2$) along the [111] direction in the pseudo-cubic representation used here.\cite{Lebeugle} 
An incommensurate antiferromagnetic order, where the spin wave has a cycloidal spiral structure with a period length of 62 nm, is etablished below the N\'eel temperature of $T_N \sim 640$~K.\cite{Smolenski,Lee, Przenios,Ismilzade,Sosnowska}
The spin wave vector propagates along the $\left[10\overline{1}\right]$ direction and lies in the $\left(\overline{1}2\overline{1}\right)$ spin rotation plane as shown in Fig. 1(a). 

Raman spectra were recorded in a backscattering geometry with a triple spectrometer Jobin Yvon T64000 using the 647.1~nm excitation line from a Ar$^+$-Kr$^+$ mixed gas laser. 
The high rejection rate of the spectrometer allows us to detect the magnons at frequencies below 50~cm$^{-1}$.
Measurements between (7 and 300~K) have been performed using an open cycle He cryostat and a LINKAM TS1500 heating stage was utilized for the 300--1000~K range. To ascertain the temperature dependence of our measurements, we have normalized the Raman spectra by the Bose factor. 
The illuminated area is about 100~$\mu$m$^2$ and the penetration depth is less than 10$^{-5}$cm.

\begin{figure}
\includegraphics*[width=7cm]{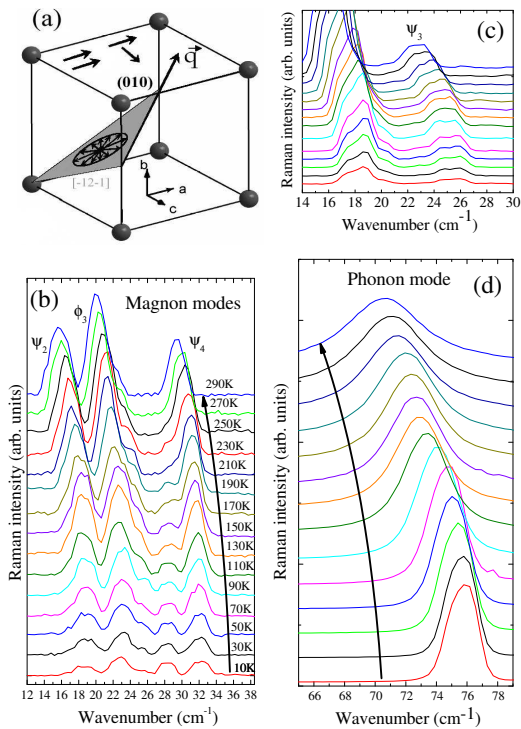}\\
\caption{\label{Figure1} (color online) 
(a) Structure of the pseudo-cubic BiFeO$_3$ unit cell showing parallel and crossed polarizations geometries in the (010) plane;
(b) parallel polarization Raman showing the spectra of the $\psi_2$, $\phi_3$ and $\psi_4$ magnons; (c) crossed polarization measurements showing the $\psi_3$ magnon; and (d) the lowest phonon mode at 75~cm$^{-1}$ measured under crossed polarization.}
\end{figure}

\begin{figure}
\includegraphics*[width=8cm]{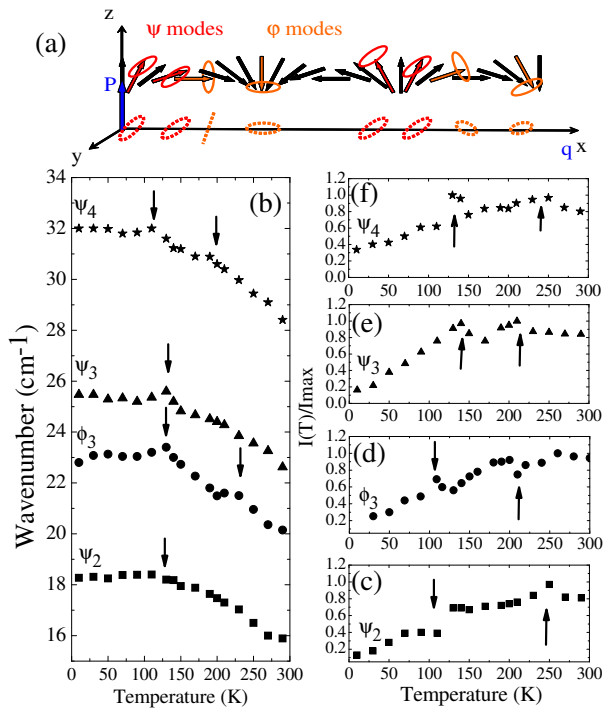}\\
\caption{\label{Figure2} (color online)
(a) Cycloidal order showing $\phi$ modes (orange ellipses) corresponding to vibrations along the tangent vector which belongs to the $xy$ plane and is perpendicular to each spin and $\psi$ modes (red ellipses) corresponding to ellipses elongated along $y$; the dashed ellipses corespond to the projection of the vibration modes in the (xy) plane (b) Temperature evolution of of the frequency of $\psi_2$ (square), $\phi_3$ (circle), $\psi_3$ (triangle), and $\psi_4$ (star) magnon modes. Normalized intensity to the maximum intensity versus temperature for  the (c) $\psi_2$ (d) $\phi_3$ (e) $\psi_3$ and (f) $\psi_4$ modes. The arrows show the anomalies discussed in the text.} 
\end{figure}

Figure \ref{Figure1} shows the magnon spectra obtained in either (b) parallel or (c) crossed polarizations, defined in the diagram of Fig. \ref{Figure1}(a). Panel (d) shows the lowest frequency $E$ phonon mode, which will be discussed later. In principle, the parallel polarization selects $\phi_n$ whereas crossed polarization gives us access to $\psi_n$ magnon modes. However, our crystal geometry does not allow to reach pure parallel polarization and a strong contribution of $\psi_2$ and $\psi_4$ modes is seen in panel (b). $\phi_n$ and $\psi_n$ modes are assigned to spin excitations in the cycloidal plane and out of this plane, respectively.  The $\phi$ mode are ellipses elongated along $y$ and the $\psi$ ellipses elongated along the tangent vector which belongs to the $xz$ plane and is perpendicular to each spin. $x$ lies along the direction of the cycloidal wavevector $\mathbf{q}$ in Fig. \ref{Figure1}(a) and $z$ is the direction of the spontaneous polarization. A diagram for $\phi$ and $\psi$ modes is shown in Fig. \ref{Figure2}(a).

Figure \ref{Figure2}(b) shows the temperature dependence of the frequencies of four magnon modes. Figures \ref{Figure2} (c)--(f) show each magnon mode intensity as a function of temperature. From Figs. \ref{Figure1}(b) and (c) and Fig. \ref{Figure2}(a) we can see that the $\psi$ and $\phi$ magnon modes shift to low energies as the temperature increases from 10~K to 300~K. This is the expected magnon behavior when one considers the thermal expansion of the lattice.
The intensity of the magnon modes, however, does not follow a conventional behavior.  Figures \ref{Figure2} (c)--(f) show that it first increases from 10~K up to 210~K before decreasing at higher temperatures. If only thermal activation was responsible for the magnon intensity, it should continuously decrease upon heating the sample.

\begin{figure}
\includegraphics*[width=8cm]{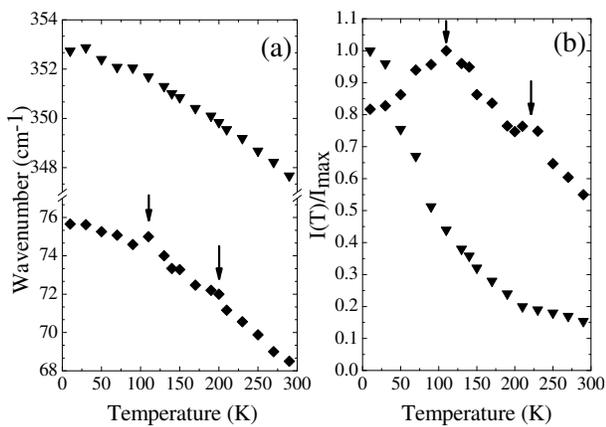}\\
\caption{\label{Figure3} Temperature dependence of (a) the frequency and (b) the normalized intensity to the maximum intensity of the lowest phonon mode (diamond). The temperature evolution of the typical phonon mode at 353~cm$^{-1}$ (triangle) is also plotted.}
\end{figure}

\begin{figure}
\includegraphics*[width=7cm]{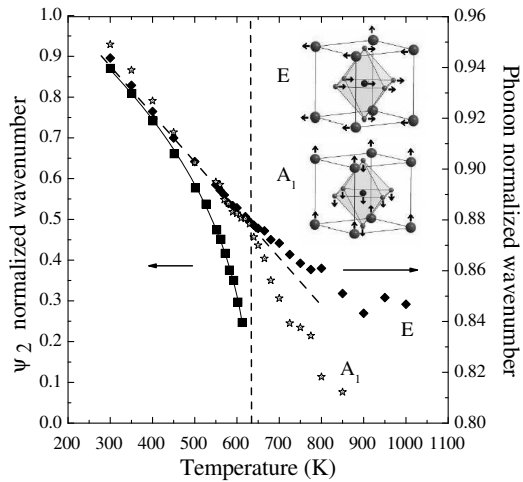}\\
\caption{\label{Figure4} High temperature dependence from 300~K to 1000~K of the frequencies of the $\psi_2$ magnon (square) and the lowest E (diamond) and A$_1$ mode (open star) normalized to their respective values at 10~K. The solid line is given by the $(1-\frac{T}{T^*})^{\alpha}$ power law. The dashed line is a guide to the eyes. The short dashed line indicates the location of $T_N$ for our samples. The inset shows the E and A1 zone-center vibrational normal modes.}
\end{figure}

Figure \ref{Figure2}(b) also shows that the slope of the $\psi_2$ frequency mode versus temperature (squares) has a small change around 125~K. The temperature dependence of the frequencies of the $\phi_3$ (circle), $\psi_3$ (triangle up) and $\psi_4$ (star) modes show small sharp peaks around 130~K and 225~K, 130~K and 110~K and 200~K, respectively.
The intensities of these modes have accidents at the same temperatures, as shown in Figs.~\ref{Figure2}(c)-(f). 

Similar anomalies in the frequencies and intensities of the magnon modes have been reported before at 130~K in Ref. \onlinecite{Cazayous1} and at 200~K in Ref. \onlinecite{Singh}. 
Here, both anomalies are observed simultaneously. The anomalies at 200K were not reported in Ref. \onlinecite{Cazayous1} because it's observation is highly sensitive to the orientation of the sample crystallographic axes compared to the polarization light vectors.

Theses anomalies have been previously interpreted as new magnetic phase transitions.\cite{Scott1} The transition around 140~K is associated to a spin reorientation but with a little coupling to the order parameter. The transition around 200~K shows a strong elastic coupling.
Although from our Raman results we cannot state that these peaks are due to novel phase transitions, a possible picture to describe them is based on spin re-orientation. The Fe$^{3+}$ magnetic moments in BFO are coupled to the oxygen atoms via the superexchange mechanism. 
A weak ferromagnetic moment results from the antisymmetric component of this interaction. 
This magnetic anisotropy is temperature dependent and leads to a spontaneous rotation of the ordered spins within a certain temperature interval called spin reorientation transition.  

Spin reorientation transitions are characterized by anomalies in the frequency and intensity of the one-magnon peak at specific temperatures. \cite{Koshizuka, Wite, Venugopalan}
In orthoferrites RFeO$_3$ (R is a rare earth), the rotation and a strong tilting of the Fe$^{3+}$ magnetic moment is made evident in Raman experiments as one needs to change the polarization as a function of temperature in order to follow the magnon signal.

In BFO, one sees a different picture. 
The anomalies in Fig.~\ref{Figure2} are observed using the same polarization configuration throughout the whole experiment. This indicates that the spin reorientations of Fe$^{3+}$ magnetic moment out of the $\left(\overline{1}2\overline{1}\right)$ cycloidal plane are rather small in comparison with the one observed in orthoferrites.

Figure \ref{Figure3} shows the temperature dependence of the (a) frequency and (b) intensity of the lowest energy $E$ phonon, introduced in Fig. \ref{Figure1}(d). Both panels of Fig. \ref{Figure3} also show, for comparison, the behavior of another $E$ phonon, located at 353~cm$^{-1}$. This latter phonon exhibits a conventional softening as the temperature is raised. The differences between the two phonons are striking. The frequency of the $E$ phonon at 75~cm$^{-1}$, associated with the motion of Bi-O bounds, exhibits a maximum at 110~K and then a shoulder close to 200~K. Its intensity increases to about 110 K before decreasing. This is the same unconventional behavior, albeit at a lower temperature, observed for magnons modes.

The accidents detected at 130 and 210~K in frequency and intensity for both the magnons and the lowest $E$ phonon mode suggest  that this phonon is coupled to the spin excitations associated with the magnetic cycloid. This gives an important hint on the magneto-electric coupling on BFO as the lowest $E$ phonon has a very strong polar character and controls the dielectric constant of the material.\cite{Lobo}

Magnon-phonon coupling is also present at high temperatures. Figure~\ref{Figure4} shows the frequency of the $\psi_2$ magnon mode at high temperatures from 300 to 625~K. The magnon softens smoothly as the temperature increases and its frequency tends to zero at $T_N$ following the $(1-\frac{T}{T^*})^{\alpha}$ power law, typical of an antiferromagnetic order parameter ($T^*$ being the temperature of the transition and $\alpha$ the order parameter exponent). The solid line in Fig. \ref{Figure4} was plotted using the parameters $T^*$=$T_N$=630~K and $\alpha$=$\frac{1}{2}$ 
expected for an antiferromagnetic transition. 
The good agreement supports the magnon mode as the antiferromagnetic order parameter.  

Figure~\ref{Figure4} also shows the thermal evolution of the normalized frequency of the lowest $E$ (75~cm$^{-1}$) and $A_1$ (144~cm$^{-1}$) modes at high temperatures.
The continuous decrease of the $E$ and $A_1$ mode frequency follows the typical variation of an optical phonon until it reaches the N\'eel temperature around 650~K.
Beyond $T_N$, the softening of the $E$ mode is reduced whereas the softening of the $A_1$ mode is enhanced. 

Although the paraelectric structure of BFO is not well established, a cubic phase at high enough temperatures is expected. The phonons in this cubic phase have a $F_{1u}$ symmetry and are triply degenerate. 
Below the Curie temperature $T_c$, the $F_{1u}$ mode is expected to split into one $E$ mode and one $A_1$ mode.\cite{Kamba}
The observation of a similar unexpected behaviour for the $E$ and $A_1$ phonons above the N\'eel temperature suggets that these two modes 
originate from the same $F_{1u}$ mode. 

Although the temperature behaviour at $T_N$ of both $E$ and $A_1$ mode show that these excitations are coupled to the spins, it is remarkable that only the $E$ mode shows low temperature anomalies coincident with these observed in the magnon response. As the ionic mouvements of the $E$ phonon are in the plane perpendicular to the polar axis, the phonon mediated coupling of the magnetic order to the ferroelectricity must involve the ferroelastic deformation of the lattice. 

In conclusion, by studing the temperature dependance of the spin excitations and the phonon modes in BiFeO$_3$ crystals, we have shown
that the lowest $E$ phonon mode is sensitive to the spin reorientation around 130 and 210~K and is 
disconnected from the magnetic sublattice when the spin cycloid disapears at $T_N$. 
Our results show that both polar phonons and magnetic excitations are sensitive to the spin order at low and high temperatures.
The symmetry of the excitations involved in the spin-phonon coupling strongly suggest that it is medaited by magneto striction and piezoelectric effects. 

This work was partially supported by the FEMMES program of the French National Research Agency.

\end{document}